\pdfoutput=1
\documentclass[twocolumn,aps,preprintnumbers,amsmath,amssymb,superscriptaddress,nofootinbib]{revtex4-1}

\usepackage{amssymb,amsmath,bbold}
\usepackage{graphicx,soul}
\usepackage[usenames,dvipsnames]{xcolor}
\usepackage[unicode=true,pdfusetitle,
 bookmarks=true,bookmarksnumbered=false,bookmarksopen=false,citecolor=Turquoise,
 breaklinks=false,pdfborder={0 0 1},backref=false,colorlinks=true,pdfpagemode=FullScreen]
 {hyperref}
\usepackage{appendix}

\begin{document}

\title{A scalar potential from gauge condensation  and its implications}

\author{Eung Jin Chun}
\email[]{ejchun@kias.re.kr}
\affiliation{School of Physics, KIAS, 85 Hoegiro, Seoul 02455, Republic of Korea}
\author{Chengcheng Han}
\email[]{hanchch@mail.sysu.edu.cn }
\affiliation{School of Physics, Sun Yat-Sen University, Guangzhou 510275, China}
\affiliation{School of Physics, KIAS, 85 Hoegiro, Seoul 02455, Republic of Korea}

\begin{abstract}
We consider a scalar field $\phi$ whose coupling to the kinetic term of a non-abelian gauge field is set at an UV scale $M$. Then the confinement of the gauge sector will induce a $\phi$-dependent vacuum energy which generates a dimensionful potential for the scalar. It provides a good example of dynamical generation of a new physics scale below $M$ through the vacuum expectation value $\langle \phi \rangle$. This mechanism may shed light on the origin of dark matter, or spontaneous symmetry breaking applicable to the electroweak symmetry.

\end{abstract}

\maketitle
\section{Introduction}
The existence of a fundamental scalar has been essential to understand an important phenomenon in the nature.
The Higgs boson, responsible for the electroweak symmetry breaking, is the only example confirmed by experiment so far.
However, it is well believed that a scalar field (inflaton) drives an early expansion of the universe~\cite{Brout:1977ix, Sato:1980yn, Guth:1980zm, Linde:1981mu, Albrecht:1982wi} to solve the flatness and horizon problems,  seeding the primordial fluctuations in cosmic microwave background~\cite{Starobinsky:1979ty, Mukhanov:1981xt}. A scalar field is also required to break Peccei-Quinn (PQ)
 symmetry at an intermediate scale through which the Strong CP conservation is enforced dynamically  \cite{Kim:2008hd}.

More recently, scalar particles have been considered extensively as a dark matter candidate for various reasons, which can also be 
extremely light~\cite{Hu:2000ke}. Furthermore, a light scalar  (dilaton) coupling to the gluon fields could lead a fifth-force between the nucleus, motivating a lot of low energy atomic experiments~\cite{Murata:2014nra}. Such a scalar field with an initial displaced vacuum 
could induce time-varying gauge couplings through its coupling to the weak and strong gauge fields, providing an first-order phase transition required by the electroweak baryogengesis~\cite{Ipek:2018lhm, Ellis:2019flb, Berger:2019yxb, Danielsson:2019ftq}.

 In this paper, we consider a scalar field $\phi$ coupling to a  general $SU(N)$ gauge field which confines at a lower scale. 
Then,  a $\phi$-dependent vacuum energy is induced and generates a new potential for the scalar boson in the confining phase.
In this way, a new energy scale can emerge to provide us some insight on understanding various phenomena mentioned above. 
The paper is organized as follows: we will first briefly overview the mechanism, and then discuss various implications in physics of
 dark matter and symmetry breaking.

\section{The mechanism}

\subsection{Confinement potential }
Let us consider an $SU(N)$ gauge theory with $n_f$ light fermions, and a singlet scalar  $\phi$ which  couples  
to the gauge field through a high dimension operator,
\begin{eqnarray}
\mathcal{L} \supset -\frac{1}{4 g^2} \left (1- c \frac{\phi}{M}  \frac {\beta} {2g}  \right ) G^{ \mu\nu} G_{\mu \nu} .
\label{formula1}
\end{eqnarray}
Note that the pre-factor  $\beta/2g$ is included to keep renormalization scale invariance of the operator, $g$ is the gauge coupling constant, and  the beta function $\beta$  is defined by  
\begin{eqnarray}
\frac{d g}{d \ln \mu} \equiv \beta=-\beta_0  \frac{g^3}{16\pi^2} 
\end{eqnarray}
with $\beta_0= (\frac{11}{3} N- \frac{2}{3} n_f)$.
If $SU(N)$ confines below $M$, there occurs gauge condensation $\langle \frac{\alpha_s}{\pi} \tilde G^{ \mu \nu} \tilde G_{\mu \nu} \rangle \simeq \Lambda^4$~\cite{Shifman:1978by, Shifman:1978bx, Gubler:2018ctz}. Here $\tilde G^{ \mu \nu} = \frac{1}{g_s} G^{ \mu \nu} $ which is the normalized  gauge field and $g_s$ should be understood as the effective gauge coupling satisfying $\frac{1}{g_s^2} =  \frac{1}{g^2}  \left (1- c \frac{\phi}{M}  \frac {\beta} {2g}  \right ) $. $\Lambda$ is the confinement scale which can be estimated as
\begin{eqnarray}
\Lambda = M \exp(-\frac{8\pi^2}{g^2\beta_0} -\frac{c}{4} \frac{\phi}{M} ) .
\end{eqnarray}
The contribution of vacuum energy from confinement is~\cite{Collins:1976yq, Pasechnik:2016sbh} \footnote{Note that a similar term is considered for the radion~\cite{vonHarling:2017yew, Fujikura:2019oyi}.},
\begin{eqnarray}
V_{vac} &=& \frac{1}{4} \langle T^\mu_\mu \rangle = \langle \frac{\beta}{8g_s} \tilde G^{ \mu \nu} \tilde G_{\mu \nu} \rangle + \frac{1}{4} (1-\gamma_m)m_f \langle \bar f f \rangle  \nonumber \\
&\simeq& - \frac{\beta_0}{32}\Lambda^4  =  - \frac{\beta_0}{32} \Lambda_0^4 \exp(- c\frac{\phi}{M} ) 
\label{potential}
\end{eqnarray}
where $\Lambda_0$ is the confinement scale when $\phi=0$ and $\gamma_m$ is the anomalous dimension of the fermion mass operator. Here we ignore the fermion contribution by assuming $m_f$ is much smaller than $\Lambda_0$.
 Since the $\Lambda$ has depends on $\phi$, it induces a potential for $\phi$. Note that this potential is only valid for $ | c \frac{\phi}{M}| \lesssim \mathcal O(1) $.  In the following we will discuss the physics implications of the emergent potential~(\ref{potential}) with the pre-factor ${\beta_0}/{32}$ absorbed by the redefined confinement scale.

\subsection{ Emergence of  a new energy scale}
As is well-known, a scale could emerge from dimension transmutation although a theory does not contain any mass parameter. 
QCD is a specific example where the condensation scale of $\Lambda_{QCD}\sim 0.1$ GeV arises due to 
the color confinement. Another example is the Coleman-Weinberg  potential which was originally used for electroweak symmetry breaking (EWSB) without introducing a dimensionful  parameter~\cite{Coleman:1973jx}.  It is also argued that such classical scale invariance may provide a solution of naturalness problem~\cite{Bardeen:1995kv}.  Along these lines, the confinement potential~(\ref{potential}),
generating a dimensionful term for the scalar $\phi$  via dimension transmutation,
 can be the source of  a new scale: the vacuum expectation value $\langle \phi\rangle$.

 Consider the following interactions of the scalar field $\phi$, 
\begin{eqnarray}
\mathcal L \supset -\frac \lambda 4 \phi^4 -\frac{1}{4 g^2} \left (1- \frac{\phi}{M} \frac{\beta}{2 g}  \right ) G^{ \mu\nu} G_{\mu \nu} 
\label{eq3}
\end{eqnarray}
which induces a potential, 
\begin{eqnarray}
V = \frac \lambda 4 \phi^4 - \Lambda_0^4 \exp(-\frac{\phi}{M} )  
\end{eqnarray}
valid for $|\phi| \lesssim M$.  From the minimization of the potential~(\ref{potential}),  fulfilling  the conditions:
$V^{\prime} =0$ and $V^{\prime \prime} > 0$ at $\phi=\langle \phi \rangle$,  
we can find the solution:
\begin{eqnarray}
&& \langle \phi  \rangle \simeq  - \lambda^{-1/3} \Lambda_0 \left ( \frac{\Lambda_0}{ M} \right )^{1/3}  \\
&& m_\phi^2 = 3 \lambda \langle \phi  \rangle^2 - \frac{\Lambda^4_0}{M^2}
\end{eqnarray}
for  $\lambda >  \frac{1}{27} (\frac{\Lambda_0}{M})^4$.

For a complex scalar field $\phi$ charged under  $U(1)$ or $Z_2$ symmetry, one can consider the Lagrangian:
\begin{eqnarray}
\mathcal L \supset -\frac \lambda 4 (\phi^\dagger \phi)^2 -\frac{1}{4 g^2} \left (1- c \frac{\phi^\dagger \phi}{M^2}  \frac{\beta}{2 g} \right ) G^{ \mu\nu} G_{\mu \nu} 
\end{eqnarray}
which induces the potential,
\begin{eqnarray}
V &=& \frac \lambda 4 (\phi^\dagger \phi)^2 - \Lambda_0^4 \exp(- c \frac{\phi^\dagger \phi}{M^2} )  \nonumber \\
&=&   \frac \lambda 4 (\phi^\dagger \phi)^2 - \Lambda_0^4 +   c \frac{\Lambda_0^4}{M^2}  \phi^\dagger \phi + \cdots .
\end{eqnarray}
In the case of $c < 0$, it provides a negative mass term and thus induces the symmetry breaking leading to
\begin{eqnarray}
&&\langle \phi  \rangle \simeq (-\frac{2 c}{\lambda})^{1/2} \frac{\Lambda_0^2}{M} \\
&& m^2_\phi \simeq -c \frac{\Lambda_0^4}{M^2} .
\end{eqnarray}

Notice that this mechanism can be applied to the PQ symmetry breaking. Taking $M$ to be the Planck scale 
$M_{P}$, one can get the axion scale of $\langle \phi \rangle \sim  10^{11}$ GeV for 
$\Lambda_0 \sim 10^{15}$ GeV with $|c|$ and $\lambda \sim \mathcal O(1)$.

\section{Dark matter and gauge condensation}

\begin{figure}[ht]
\centering
\includegraphics[width=2.5in]{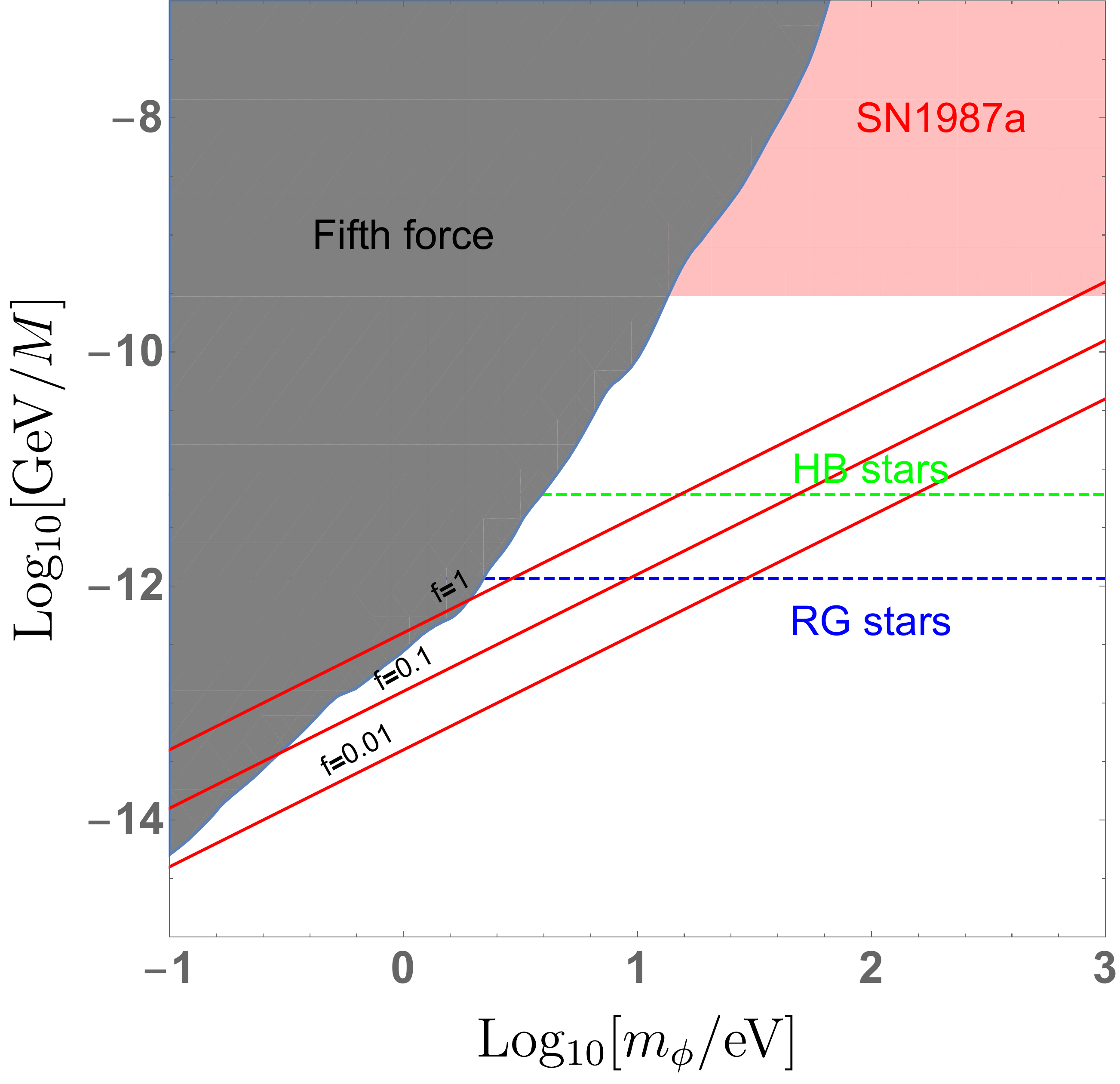}
\caption{The parameter space where $\phi$ as a dark matter candidate. We show the limit from fifth-force measurement \cite{Murata:2014nra} (gray region) and stellar cooling limits from HB stars \cite{Hardy:2016kme} (above green curve), RG stars \cite{Hardy:2016kme} (above blue curve) and SN1987A \cite{Knapen:2017xzo} (pink region). The red curve is the parameter space where $\phi$ satisfies the dark matter relic density with a fraction $f$.
}
\label{fig1}
\end{figure}

\subsection{Minimal model from color confinement}
Consider a scalar dark matter $\phi$ which  couples only to the gluon field. In this case, its abundance can be produced through the misalignment  of the vacuum before and after the QCD  phase transition. Assuming again the following interactions,
\begin{eqnarray}
\mathcal L \supset -\frac \lambda 4 \phi^4 -\frac{1}{4 g_s^2} \left (1- \frac{\phi}{M}   \frac {\beta_s} {2g_s}  \right ) G_s^{a \mu\nu} G_{s\mu\nu}^a
\label{qcd}
\end{eqnarray}
where we used  the index $s$ to denote the color $SU(3)_c$ sector.
Before QCD confines,  the  $\frac 1 4 \lambda \phi^4$ dominate the potential having the minimum at $\langle \phi  \rangle =0$.  After the confinement of QCD,  the confinement potential arises
\begin{eqnarray}
V = \frac \lambda 4 \phi^4 -  \Lambda^4_{\rm QCD} \exp(- \frac{\phi}{M} )
\end{eqnarray}
developing a new minimum at 
\begin{equation}
\langle \phi  \rangle \simeq  -  \lambda^{-1/3} \Lambda_{\rm QCD} 
\left ( \frac{\Lambda_{\rm QCD}}{ M} \right )^{1/3}
\end{equation}
where  we  define $\Lambda^4_{\rm QCD} \equiv \frac{9}{32} \langle \frac{\alpha}{\pi}G^{ a \mu \nu}G^a_{\mu \nu} \rangle$ $={9\over 32} 0.028$ GeV$^4$~\cite{Horsley:2012ra} \footnote{Note that there is a large uncertainty in the determination of  $\langle \frac{\alpha}{\pi}G^{a \mu \nu} G^a_{\mu \nu} \rangle$~\cite{Narison:2018dcr}.}.
At the same time, $\phi$ gets the mass $m_\phi \simeq \sqrt{3\lambda} \langle \phi  \rangle $ which is much larger than the Hubble parameter. Thus, $\phi$ starts immediately to oscillate around the new minimum. If $\phi$ contributes all the dark matter component, the relic density of $\phi$ should satisfy
\begin{eqnarray}
\rho_{\phi} \times \left ( \frac{ T_{eq}}{T_{osc}} \right )^3 \frac{ g_{*} (T_{eq})}{g_{*} (T_{osc} ) } \simeq 0.4 ~\rm (eV)^4
\label{darkmatter}
\end{eqnarray}
where $\rho_{\phi} \simeq \frac{3}{4}{\lambda} \langle \phi  \rangle^4$ and $T_{eq} \simeq 0.8$ eV is the temperature at the matter-radiation equality. The oscillation temperature is taken to be  $T_{osc} = 0.15 $ GeV~\cite{Petreczky:2012rq}.

In our scenario the dark matter coupling to photons arises  at two loop level,  The decay width of the light scalar can be estimated as,
\begin{eqnarray}
\Gamma_{\phi\to \gamma\gamma} \approx \sum_q  \frac{1}{\pi} \left (\frac{\alpha}{4\pi} \right)^2  \alpha_s^2 Q_q^4 \frac{m_\phi^3}{M^2} . 
\end{eqnarray}
Where $Q_q$ is the electric charge of the quarks. The minimization conditions and the dark matter relic density relation (\ref{darkmatter}) set the dark matter lifetime as a function of its mass and fraction $f$:
\begin{equation}
\tau_{\phi \to \gamma \gamma} \approx  5 \times 10^{17}  \mbox{sec}\, 
\left ( \frac{1}{f} \right ) \left( \mbox{3 keV} \over m_\phi \right)^5 .
\end{equation} 
Thus, the dark matter mass should not be larger than about 3 keV.

On the other hand, the interaction in Eq.~(\ref{qcd}) also induces a coupling between the scalar and nucleus at low energy:
\begin{eqnarray}
\frac{\phi}{M} \frac{\beta}{2g} G^{a \mu\nu} G^a_{\mu \nu}  \Rightarrow \mathcal O(1) \frac{\phi}{M} m_N \bar N N .
\end{eqnarray}
The fifth-force measurement and  the astrophysical observations set a strong limit on such couplings.  In Fig.~\ref{fig1} we show the dark matter preferred region as well as these limits. The red curve is the parameter space where $\phi$ satisfies the dark matter relic density. It shows that this scalar can hardly provide all the the dark matter component surviving all the constraints.  
More parameter space is available if we consider $\phi$ contributes to a fraction of the dark matter.

\subsection{Additional  SU(N) potential}
Let us now generalize the previous consideration by adding an additional confining gauge sector of $SU(N)$:
\begin{eqnarray}
\mathcal{L} \supset && - \frac{1}{4} \lambda \phi^4 -\frac{1}{4 g_N^2} \left (1-  \frac{\phi}{M} \frac{\beta_N}{2g_N}\right ) G^2_N  \nonumber \\
&& -\frac{1}{4 g_s^2} \left (1-  \frac{\phi}{M} \frac{\beta_s}{2g_s}\right ) G^2_s
\end{eqnarray}
where we assumed the same coupling of $\phi$ to the $SU(N)$ and $SU(3)_c$ gauge fields for simplicity. If we allow smaller coupling of $\phi$ to the QCD sector, our bounds will be relaxed accordingly. 
The above interactions can be obtained after integrating out by a heavy fermion in bi-fundamental representations of $SU(N)$ and $SU(3)_c$. Then the total confinement potential at low energy becomes, 
\begin{eqnarray}
V &=&   \frac{1}{4} \lambda \phi^4 - \Lambda_{N}^4 \exp(-{\phi \over M}) 
- \Lambda_{\rm QCD}^4 \exp(-{\phi \over M}) .
\end{eqnarray}
Assuming $\Lambda_{N} \gg \Lambda_{\rm QCD}$ \footnote{Note that the $\Lambda_{\rm QCD}$ here should be changed into $\Lambda^\prime_{\rm QCD}=\Lambda_{\rm QCD} \exp(\frac{1}{4}\frac{\langle \phi \rangle_{\rm min}}{M})$ since the gauge coupling get renormalized for $\phi \ne 0 $. However, for our case  $M \gg \langle \phi \rangle_{\rm min}$, this difference is very small and therefore we safely neglect it. Same argument is for the pure QCD case. }, the  scalar potential is dominated by the $SU(N)$ confinement leading to the previous relations (7,8) with $\Lambda_0 \to \Lambda_N$.
The requirement for the dark matter relic density  (16) is again applicable with  $T_{osc}=\Lambda_N$.
To maintain the two sectors in thermal equilibrium before the hidden sector confines we may add more fermions with smaller mass. 

\begin{figure}[ht]
\centering
\includegraphics[width=2.5in]{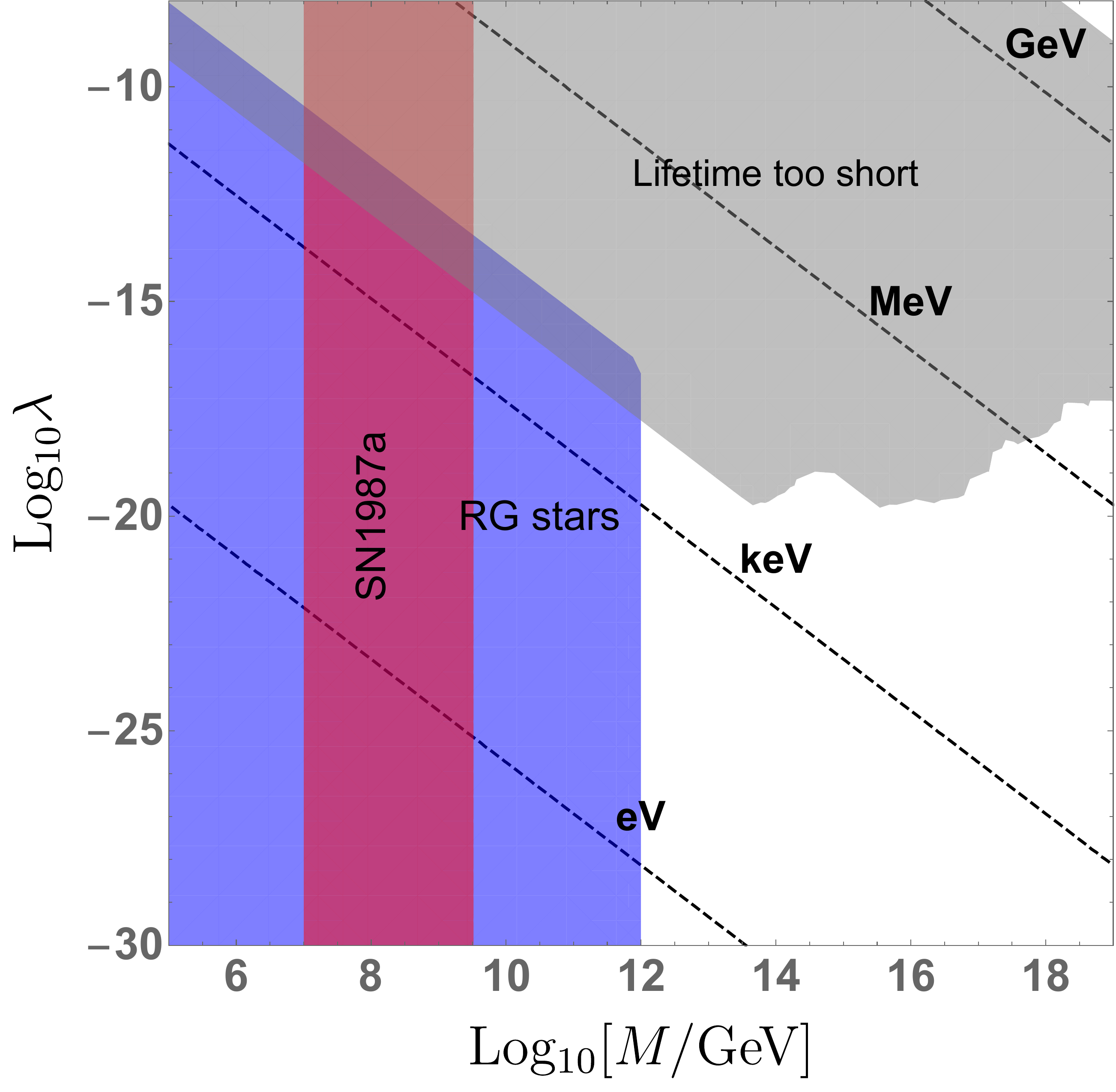}
\caption{The parameter space where $\phi$ as a dark matter candidate. All of shaded region are excluded by different searches. The dashed curves show $m_\phi =$ 1 eV, 1 keV, 1 MeV, 1 GeV respectively.
}
\label{fig2}
\end{figure}

In Fig.~{\ref{fig2}} we show the viable parameter space.  The dashed curves show the values of $m_\phi$.  The gray region is excluded by the X-ray or $\gamma$-ray searches  due to the decaying of the scalar into photons~\cite{Essig:2013goa}. These limits generally require the lifetime of scalar to be larger than $10^{27}$ seconds.  
In the present case, the dark matter mass can be in the range of (eV, MeV) where the upper limit is set by the condition:
$M < M_P$ as a reference. 

\section{ Origin of the Electroweak symmetry breaking}

It is known that the electroweak symmetry breaking is triggered by the condensation of Higgs boson. Although  the Higgs properties of the standard model are well established, there still  to be understood:  "What is the origin of the Higgs potential?".
 Here we attempt to drive the electroweak symmetry breaking by applying our mechanism of dimensional transmutation.  

\begin{figure}[ht]
\centering
\includegraphics[width=2.7in]{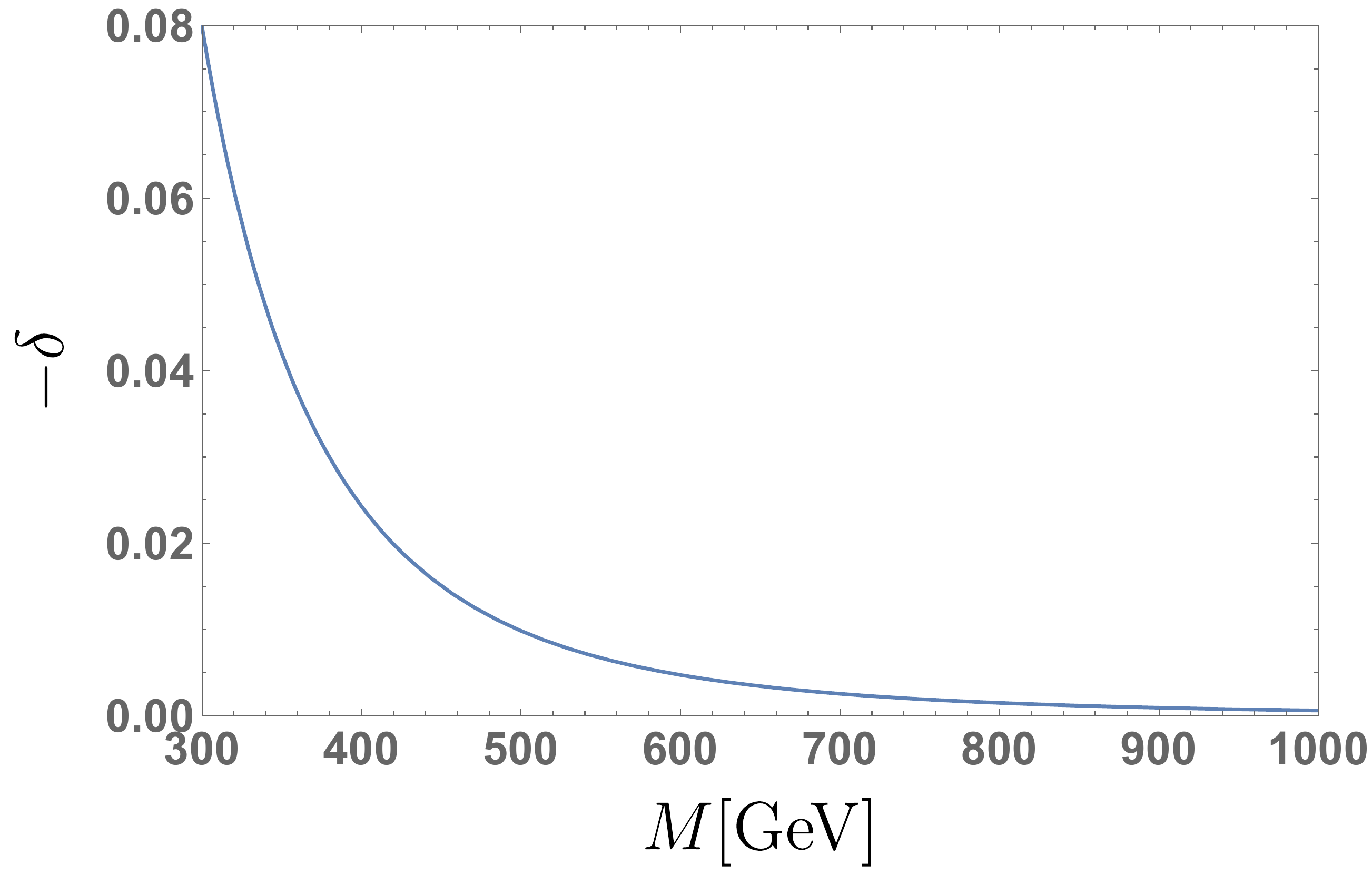}
\caption{ Deviation of the Higgs trilinear coupling for different  $M$. }
\label{fig3}
\end{figure}

Assuming that the Higgs couples to a hidden strong sector through a dimension-6 operator:  
\begin{eqnarray}
\mathcal L \supset - \lambda  (H^\dagger H)^2 -\frac{1}{4 g^2} \left (1+ \frac{H^\dagger H}{M^2}  \frac{\beta}{2 g} \right ) G^{ \mu\nu} G_{\mu \nu}  \nonumber \\
\end{eqnarray}
we get the  Higgs potential
\begin{eqnarray}
V &=&  {\lambda} (H^\dagger  H)^2 -  \Lambda_0^4 \exp( \frac{H^\dagger  H}{M^2} )   \nonumber   \\
 &\approx&  -\frac{\Lambda_0^4}{ M^2} H^\dagger H 
 +  (\lambda+\frac{1}{2} \frac{H^\dagger  H}{M^2}   ) (H^\dagger  H)^2    \nonumber \\
&& -\frac{1 }{6} \frac{\Lambda_0^4}{ M^6} (H^\dagger  H)^3    + \cdots . 
\end{eqnarray}
Taking   $ H =  ( 0 , (h+v)/\sqrt2 )^T $ in the unitary gauge, the minimization conditions give us 
\begin{eqnarray}
\Lambda_0 = \sqrt 2 M  \left (\frac{m_h^2}{8 M^2- v^4/M^2}  \right )^{1/4}
\end{eqnarray}
where $v=246$ GeV and $m_h=125$ GeV. 
The higher dimension operators modify the Higgs trilinear coupling $\lambda_{hhh}$. Defining $\delta \equiv (\lambda_{hhh}/\lambda^{SM}_{hhh})-1$, we find 
\begin{equation}
\delta = - \frac{4 v^4}{ 24 M^4-3 v^4} \simeq  -\frac{1}{6} (\frac{v}{M})^4
\end{equation}
which is highly suppressed. 
As can be seen in Fig.~\ref{fig3}, the deviation of Higgs trilinear coupling from the standard model prediction can be sizable  for 
$M= 300-400$ GeV, which might be probed at future colliders. It drops quickly below 1\% for  $M> 500$ GeV.
For this range, we need $\Lambda_0 \gtrsim 200$ GeV.

\section{{Conclusion}}
We studied a new mechanism where  a scalar potential emerges from the scalar coupling to a $SU(N)$ gauge field strength which confines at a low energy scale. Such a potential may play an important role to understand the origin of dark matter, or spontaneous breaking of symmetry such as PQ and  electroweak symmetry. 
It will be an interesting task to see whether the potential for the inflaton or quintessence can emerge in a same way.  
\\

\noindent {\bf{Acknowledgements}}
This project has received support from the European Union’s Horizon 2020 research and innovation programme under the Marie Skłodowska-Curie grant agreement No 690575. E.J.C. acknowledges support from InvisiblesPlus RISE No. 690575. C. H thanks Shi Pi for helpful discussions. CH acknowledges support from the Sun Yat-Sen University Science Foundation.

\end{document}